\documentclass[11pt]{article}

\usepackage[margin=1in]{geometry}
\usepackage{amsmath,amssymb,mathtools}
\usepackage{siunitx}
\usepackage{graphicx}
\usepackage{subcaption}
\usepackage{float}
\usepackage{placeins}
\usepackage{xcolor}
\usepackage[hidelinks]{hyperref}
\hypersetup{hidelinks, pdfborder={0 0 0}, colorlinks=false}
\usepackage{cleveref}
\usepackage{booktabs}
\usepackage{enumitem}

\graphicspath{{./}{./fig/}{./figs/}{./Figures/}{./images/}}
\DeclareGraphicsExtensions{.png}

\newcommand{\IfFigExists}[3]{%
  \IfFileExists{#1.png}{#2}{%
    \IfFileExists{figs/#1.png}{#2}{%
      \IfFileExists{fig/#1.png}{#2}{%
        \IfFileExists{Figures/#1.png}{#2}{%
          \IfFileExists{images/#1.png}{#2}{#3}%
        }%
      }%
    }%
  }%
}

\makeatletter
\newcommand{\incfig}{\@ifnextchar[\incfig@opt\incfig@noopt}
\newcommand{\incfig@opt}[3][]{\incfig@noopt{#1}{#2}{#3}}
\newcommand{\incfig@noopt}[3]{%
  \IfFigExists{#1}{\includegraphics[#2]{#1}}{%
    \fbox{\parbox{0.96\linewidth}{\vspace{2.6cm}\centering\textbf{Missing figure:} \texttt{\detokenize{#1}.png}\\[2pt]#3\vspace{2.6cm}}}%
  }%
}
\makeatother

\newcommand{\FigSetupLamp}{setup_final}          
\newcommand{\FigChopper}{chopper_framerate}     
\newcommand{\FigArcDemo}{arc_application}       

\title{Passive Incoherent Ultrafast Mid-Infrared Upconversion Imaging and Its Calibration}

\author{\parbox{\textwidth}{\centering
\normalsize
Jin-Peng Li$^{1,2,3}$, Zhi-You Li$^{1,2,3}$, Zhao-Qi-Zhi Han$^{1,2,3}$, Xiao-Hua Wang$^{1,2,3}$, He Zhang$^{1,2,3}$, Yin-Hai Li$^{1,2,3,4}$, Bo-Wen Liu$^{1,2,3}$,Wen-Tao Luo$^{5}$ , Zhi-Yuan Zhou$^{1,2,3,4,*}$ ,and Bao-Sen Shi$^{1,2,3,*}$\\
\small
$^{1}$Laboratory of Quantum Information, University of Science and Technology of China, Hefei 230026, China;\\
\small
$^{2}$CAS Center for Excellence in Quantum Information and Quantum Physics, University of Science and Technology of China, Hefei 230026, China;\\
\small
$^{3}$Anhui Province Key Laboratory of Quantum Network, University of Science and Technology of China, Hefei 230026, China;\\
\small
$^{4}$Anhui Kunteng Quantum Technology Co. Ltd., Hefei 231115, China\\
\small
$^{5}$National Key Laboratory of Deep Space Exploration, Hefei 230026, China\\
\small
}}

\date{}

\usepackage[backend=biber,style=numeric,sorting=none,maxnames=3,minnames=3,giveninits=true,natbib=true]{biblatex}
\begin{document}
\maketitle
\begin{abstract}
Ultrafast mid-infrared (MIR) imaging is a key enabling capability for monitoring transient thermal and plasma phenomena in scientific diagnostics and industrial safety. However, conventional cryogenic MIR cameras face a fundamental trade-off between frame rate, noise, and pixel format. Here we report a passive, incoherent MIR imaging platform that leverages sum-frequency upconversion in chirped periodically poled lithium niobate (CPLN) to translate broadband 3--5\,$\mu$m scenes to the near-infrared, enabling ultrafast acquisition on a silicon-based intensified CCD (iCCD). In fast-kinetics mode we achieve a physical frame rate of 100\,kHz with microsecond-scale gate control, and we directly capture the full evolution of an air-breakdown electric arc, resolving its rapid ignition, expansion, and decay dynamics. Beyond demonstrating ultrafast passive imaging, we introduce a drift-aware calibration workflow based on Allan deviation analysis to quantitatively select the gate width and averaging strategy under realistic slow-drift and multiplicative noise. This combined capability---ultrafast passive MIR imaging plus operationally meaningful calibration---provides a practical route toward real-time thermal surveillance and early-warning systems for hazardous fast transients.
\end{abstract}

\vspace{0.5em}

\section{Introduction}
\label{sec:intro}

High-speed mid-infrared (MIR) imaging provides direct access to temperature fields and thermal-radiation dynamics in a wide range of scenarios, including combustion diagnostics, high-voltage fault monitoring, material processing, and remote thermal surveillance. In many engineering contexts, the most critical requirement is not merely to \emph{detect} a hazardous event, but to \emph{issue an early warning} by capturing the fast precursors and the onset phase with sufficient temporal resolution. Electric arcs and flashovers exemplify such transients: the energy deposition, plasma expansion, and radiative output can evolve on microsecond-to-millisecond time scales, leaving a narrow window for prediction, mitigation, or system shutdown.

In practice, most transient thermal events such as electric arcs is often detected with ultraviolet (UV) sensors or cameras, leveraging their strong sensitivity to discharge emission and fast response. However, UV-based monitoring is frequently deployed in an event-triggering mode: it excels at confirming that an arc has occurred, yet it provides limited access to the spatiotemporal \emph{precursors} that could enable actionable early warning\cite{Park2024ArcFlashUV}. Equally important, the UV approach is commonly optimized for local, near-range deployment---for example, within enclosures or at fixed viewpoints where background rejection and line-of-sight constraints can be tightly controlled. Extending UV monitoring to stand-off scenarios typically requires more stringent optical alignment, shielding, and site-specific configuration, and it can become less convenient to engineer as a general, easily deployable solution for wide-area surveillance\cite{Riba2022UVImagingDischargeReview}. By contrast, MIR thermal emission directly reflects the underlying heating and expansion dynamics, and it resides in atmospheric transmission windows, making remote and wide-area monitoring more practical in open environments\cite{Shen2023TransportableMIRLHR}. These considerations motivate an ultrafast MIR imaging capability that can resolve the onset evolution and support physically grounded decision variables for early warning, while remaining compatible with passive, stand-off sensing.

Despite the relevance of ultrafast MIR imaging, conventional MIR focal-plane arrays (e.g., HgCdTe/InSb) typically require cryogenic cooling and are constrained by readout bandwidth and sensitivity at high frame rates. Frequency upconversion imaging offers an alternative route by nonlinearly translating MIR photons to shorter wavelengths compatible with mature silicon detectors. Upconversion MIR imaging has progressed rapidly, from early demonstrations of room-temperature upconversion-based thermal imaging \cite{Dam2012NP} to video-rate MIR upconversion systems \cite{Junaid2019Optica}. More recently, wide-field single-photon MIR upconversion imaging has enabled ultrafast imaging by combining active illumination with tightly synchronized pump gating\cite{Huang2022NatCommun}. Such schemes typically require active illumination together with tight pump--signal timing synchronization, which can complicate deployment and constrain their applicability to passive, incoherent scenarios. In addition to the works cited above, some foundational studies established the incoherent-image upconversion theory and high-resolution 2D upconversion of incoherent light \cite{Dam2012OE,Dam2010OL}. Related efforts have further clarified performance metrics such as NETD and illumination-dependent fidelity in upconversion thermal cameras \cite{Ge2023APN_NETD,Ge2023PRApplied}.

Beyond upconversion-based approaches, a number of complementary strategies have also been explored to accelerate MIR acquisition, including photothermal and interferometric modalities \cite{Tamamitsu2020Optica,Paiva2022AnalChem} and quantum-enabled paradigms such as microscopy with undetected photons \cite{Kviatkovsky2020SciAdv,Lahiri2015PRA}. Recent progress has also demonstrated how two-photon absorption imaging can be combined with fast spectral filtering and agile tuning to deliver hyperspectral videography \cite{Fang2023NatCommun}. Together, these directions underscore a broad need for fast MIR acquisition, yet they often involve specialized excitation/interferometric stability or wavelength-agile control that can be nontrivial to engineer for passive stand-off monitoring. This context motivates the present work focused on passive, incoherent ultrafast imaging and its calibration under realistic drift, with an emphasis on deployment-oriented early warning and stand-off monitoring of hazardous transients.

Notwithstanding these advances, two practical gaps remain for deploying upconversion imagers in engineering early-warning tasks. First, many high-speed demonstrations rely on specialized pump modulation or multiplexing schemes; in contrast, a passive surveillance scenario often involves \emph{incoherent} broadband thermal emission from uncontrolled sources, for which the system must operate robustly under drift and fluctuating backgrounds. Second, the operating point (gate width, averaging strategy, ROI definition) is frequently chosen heuristically, yet early-warning performance is ultimately limited by a balance between additive noise (e.g., background and readout) and multiplicative noise (e.g., pump fluctuations and slow drift). This gap is particularly relevant when the goal shifts from ``event confirmation'' to ``precursor-aware warning'': reliable early warning hinges on extracting subtle growth and fluctuation signatures prior to the bright phase, which makes drift-aware calibration and statistically robust decision variables essential.

In this work, we focus on a deployment-oriented capability: passive incoherent MIR upconversion imaging with ultrafast acquisition, demonstrated on the full evolution of an air-breakdown electric arc at 100\,kHz. In this framing, the arc is not merely a vivid target but a representative fast transient for which resolving onset dynamics opens a practical warning window: the images allows one to characterize growth rates, expansion dynamics, and fluctuation signatures in the pre-bright regime that would be difficult to access with slower imagers or threshold-triggered sensing. Importantly, we reposition Allan deviation analysis as a \emph{calibration tool} that supports the primary imaging capability: it provides a quantitative, drift-aware criterion to select gate width and averaging time in the presence of realistic noise.

Our paper is therefore structured to (i) establish the system concept and experimental design of passive, incoherent MIR-to-VIS/NIR upconversion imaging, (ii) validate the ultrafast capability through dedicated performance characterization and a 100\,kHz arc-dynamics demonstration framed by an early-warning motivation, and (iii) introduce drift-aware calibration and practical operating guidelines based on Allan deviation analysis, which together translate high-frame-rate imaging into quantitatively reliable, deployment-oriented sensing.

\section{System setup and experimental methods}
\label{sec:setup}

\subsection{System setup}
\label{sec:setup_arch}

\begin{figure}[t]
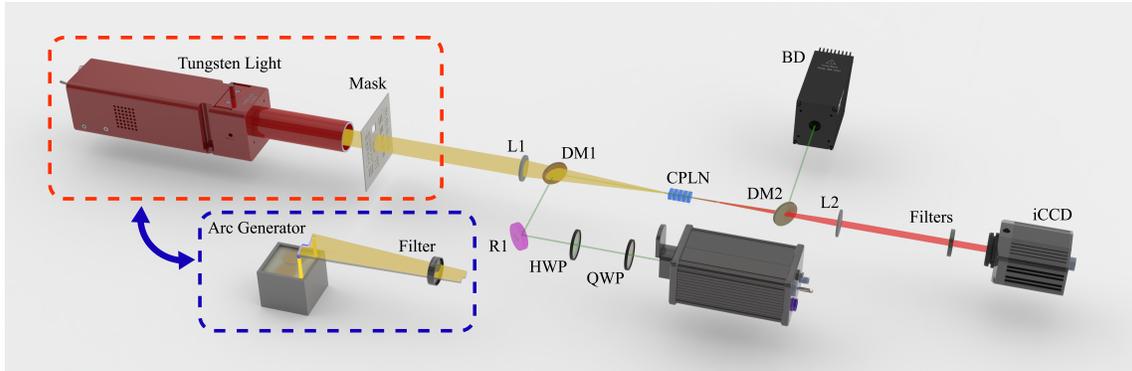

  \centering
  \incfig{\FigSetupLamp}{width=0.92\linewidth}{System optical layout for incoherent characterization with tungsten-lamp illumination.}
  \caption{System optical layout for passive incoherent MIR upconversion imaging. Mask, resolution target; L1--L2, MIR and VIS/NIR lenses; DM1--DM2, dichroic mirrors; HWP/QWP, half-/quarter-wave plates; R1, high-reflect (HR) mirror; CPLN, periodically poled congruent lithium niobate; BD, beam dump; iCCD, intensified charge-coupled device.}
  \label{fig:setup_lamp}
\end{figure}

Figure~\ref{fig:setup_lamp} illustrates the optical layout of our passive incoherent MIR upconversion imaging system. A broadband incoherent MIR scene is collected and combined with a continuous-wave 1064\,nm pump, and both are focused into a CPLN crystal to realize sum-frequency generation (SFG). The upconverted near-infrared image is then relayed to a silicon iCCD for gated acquisition. This layout serves as a unified hardware baseline for all experiments reported in this work. For controlled calibration and performance characterization, the MIR scene is provided by a tungsten--halogen lamp illuminating a transmissive resolution target. For arc-dynamics measurements, the scene-side source is replaced by an electric arc while the pump path, upconversion stage, filtering, and detection chain remain unchanged, enabling consistent comparison between controlled tests and ultrafast passive monitoring.

In the tungsten-lamp configuration (Fig.~\ref{fig:setup_lamp}), a tungsten--halogen light source provides broadband incoherent MIR radiation with a built-in filter ($\lambda_0=\mathrm{4000\,nm}$, $\Delta\lambda=\mathrm{2000\,nm}$) to illuminate a transmissive resolution target (mask). The transmitted MIR is collected by a CaF$_2$ lens (L1, $f=\mathrm{150\,mm}$) and overlapped with a continuous-wave 1064\,nm pump in the CPLN crystal. After upconversion, dichroic mirrors (DM1--DM2) remove the residual pump; the rejected 1064\,nm beam is routed into a beam dump (BD) for safe absorption. The upconverted light is collected by a B-band lens (L2, $f=\mathrm{150\,mm}$), filtered by a stack comprising two 700\,nm long-pass and three 900\,nm short-pass filters to suppress residual pump leakage and second-harmonic components, and then imaged onto the iCCD sensor.

The upconversion process follows energy conservation $\omega_u=\omega_s+\omega_p$ and quasi-phase matching in CPLN. In incoherent operation, the converted signal is effectively integrated over the admitted emission bandwidth set jointly by the phase-matching acceptance and the spectral filtering. Consequently, the conversion efficiency and noise floor are strongly influenced by pump-intensity stability, crystal temperature stability, and stray-light suppression. We use a $\mathrm{3\times2\times30\,mm^3}$ CPLN crystal with a chirped poling period spanning 21.6--23.4\,$\mu$m, pumped by a 30\,W continuous-wave laser at 1064\,nm. The pump beam is maintained at an approximately $\sim$1\,mm diameter in the crystal to support high power while minimizing aperture effects. The CPLN temperature is stabilized at $\mathrm{23.3\,^{\circ}C}$; in the experiment, a precision temperature controller keeps fluctuations within $\mathrm{2\,mK}$ to prevent phase mismatch during prolonged pumping. For short-duration measurements the crystal can operate near room temperature, whereas active stabilization becomes beneficial when long records are acquired for stability analysis.

Arc imaging reuses the same upconversion and detection chain, but replaces the lamp/target scene by an electric-arc source at the scene side. The arc emission is first filtered by a MIR bandpass ($\lambda_0=\mathrm{4000\,nm}$, $\Delta\lambda=\mathrm{2000\,nm}$) to suppress VIS/NIR radiation from directly entering the system, and is then injected into the same collection path toward L1 and the CPLN stage. This ``source replacement'' strategy ensures that any changes observed between controlled calibration and arc-dynamics experiments originate primarily from the scene dynamics rather than from modifications of the upconversion optics.

\subsection{Experimental methods}
\label{sec:methods}

Ultrafast image sequences are acquired using an intensified CCD (Andor DH334T) operated in fast-kinetics mode. In this mode a sequence of masked storage rows enables rapid acquisition of a burst of frames at a fixed physical frame interval; with a reduced region of interest, the physical frame rate can reach 100\,kHz (10\,$\mu$s frame interval). The intensifier provides gated integration with a user-defined gate width $\tau_g$, which determines the effective integration time per frame and sets the trade-off between photon budget and saturation margin. In this work, four representative gate widths $\tau_g\in\{10,20,50,100\}\,\mu$s are used specifically for Allan calibration to quantify stability and drift behavior as a function of per-frame integration.

Allan calibration is performed under the tungsten-lamp configuration in Fig.~\ref{fig:setup_lamp} using only the transmissive target as the scene. For each gate width, we record long time series of 2000 frames and compute ROI-based metrics frame by frame. Allan deviation is then evaluated versus averaging time to identify the transition from white-noise-dominated averaging to drift-limited behavior, yielding a drift-aware optimal integration horizon and an empirical uncertainty bound that is directly relevant to deployment.

To validate the ultrafast temporal sampling capability in a visually intuitive manner, we perform a rotating-chopper experiment using the same tungsten-lamp illuminated target. A rotating chopper blade is positioned adjacent to (and effectively ``attached'' to) the transmissive region of the target so that its rotation progressively occludes the transmitted MIR field. Capturing the continuous occlusion of the target stripes at a 10\,$\mu$s frame interval provides a direct demonstration of true 100\,kHz passive imaging without relying on assumptions about periodicity or external synchronization.

For arc-dynamics measurements, the lamp and target are removed and an electric arc is used as the passive MIR source. The iCCD is operated at 100\,kHz continuous acquisition setting to record the arc evolution, enabling observation of rapid transient stages that are relevant to early-warning scenarios.

Throughout the paper we quantify system response using ROI-based metrics. Let $I_k(\mathbf{r})$ denote the pixel intensity in frame $k$ at spatial coordinate $\mathbf{r}$. For an ROI $\Omega$, we define the mean intensity
\begin{equation}
\label{eq:metric_mean}
\mu_k = \frac{1}{|\Omega|}\sum_{\mathbf{r}\in\Omega} I_k(\mathbf{r}),
\end{equation}
and, when a bright region $\Omega_b$ and a nearby dark/background region $\Omega_d$ are both available, the differential metric
\begin{equation}
\label{eq:metric_diff}
\Delta_k = \frac{1}{|\Omega_b|}\sum_{\mathbf{r}\in\Omega_b} I_k(\mathbf{r}) - \frac{1}{|\Omega_d|}\sum_{\mathbf{r}\in\Omega_d} I_k(\mathbf{r}).
\end{equation}
The differential metric suppresses additive offsets from stray light and camera bias, which is particularly important for long-term stability and for early-warning decisions based on small signal increases.

\section{Results}
\label{sec:results}

\subsection{Ultrafast MIR imaging}
\label{sec:results_ultrafast}

We first establish the ultrafast capability of the complete passive upconversion-imaging chain (optics + upconverter + iCCD readout) under controlled incoherent illumination, and then demonstrate ultrafast passive imaging of a representative hazardous transient. This ``validation--demonstration'' sequence ensures that the observed dynamics are interpreted against an experimentally verified temporal sampling baseline.

\begin{figure}[t]
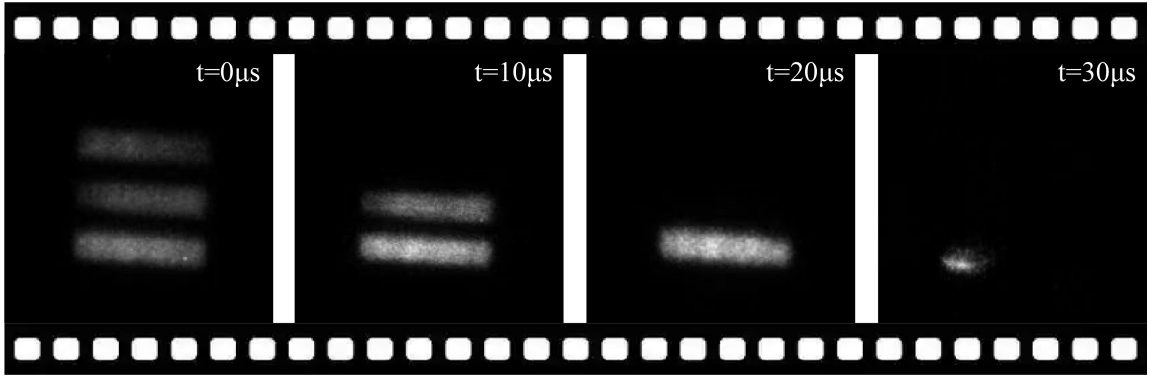

  \centering
  \incfig{\FigChopper}{width=0.92\linewidth}{Rotating-chopper measurement used to validate effective frame rate.}
  \caption{Frame-rate verification using a rotating chopper under incoherent illumination. The known modulation is recovered in the recorded high-speed burst, confirming the physical 100\,kHz sampling of the system.}
  \label{fig:chopper}
\end{figure}

To validate the effective temporal sampling, we modulate the incoming incoherent MIR scene using a rotating chopper wheel. The chopper imposes a known temporal modulation that maps to the recorded frames, allowing verification of the physical frame interval and potential aliasing artifacts. Figure~\ref{fig:chopper} shows representative chopper-recorded data. In our implementation, the system records a burst at a physical frame rate of 100\,kHz (10\,$\mu$s frame interval) while the chopper is operated at 10\,kHz. The chopper wheel used in this experiment contains 100 equally spaced slots, such that a rotation rate of 100\,Hz corresponds to a 10\,kHz intensity modulation. The chopper blade used for occlusion has an effective radius of $\sim\mathrm{5\,cm}$ and rotates at $\mathrm{100\,Hz}$, corresponding to a tangential speed of $\mathrm{31.4\,m/s}$ as it sweeps across the transmissive region of the resolution target. The transmissive aperture of the target is approximately $\mathrm{1\times1\,mm^2}$ and evolves from fully transmissive to nearly fully blocked after three exposure cycles. This behavior is in good agreement with the chopper rotation speed, evidencing the system's high-frame-rate imaging capability and its ability to resolve rapidly moving objects under incoherent illumination.

In this test, the tungsten-lamp illuminated resolution target provides a spatially structured, quasi-static scene, while a rotating mechanical chopper is placed adjacent to the transmissive window of the target. As the blade sweeps across the window, the transmitted MIR flux is progressively occluded, converting the spatially fixed stripe pattern into a well-defined temporal modulation at the camera. Recording this gradual occlusion process at 100\,kHz therefore provides a direct, system-level validation of the attainable frame rate under passive, incoherent illumination, following the strategy adopted in prior ultrafast upconversion imaging.

Physically, the chopper converts a stationary spatial pattern into a deterministic time sequence: each pixel experiences an approximately step-like transmission as the blade edge passes. At 100\,kHz sampling, the measured transition width in frame counts directly reflects the blade-edge velocity projected onto the object plane and the optical magnification. Accordingly, the absence of temporal smearing at the edge and the recovery of the expected periodicity indicate that the effective integration time is short compared with the modulation period, and that no additional timing bottleneck is introduced by the upconversion stage or the readout pipeline. Because the camera samples at 100\,kHz (10\,$\mu$s frame interval), a 10\,kHz chopper frequency corresponds to a 100\,$\mu$s period, i.e., ten frames per modulation cycle. This one-to-ten mapping is evident in the recovered waveform and provides a direct cross-check that the temporal sampling is governed by the physical readout cadence rather than an effective, down-sampled rate. The finite gate width broadens the transitions at the edges of each bright segment, which is expected because each frame integrates over a nonzero exposure window; nevertheless, the modulation contrast remains high, indicating that the upconversion stage and the subsequent filtering do not smear the temporal response within the gate. Taken together, this measurement verifies that the system can faithfully record intensity dynamics on the 10\,$\mu$s time scale under incoherent illumination.

Having established the system-level temporal sampling, we next demonstrate ultrafast passive MIR imaging on an air-breakdown electric arc. To contextualize the observed spatiotemporal evolution, we note that atmospheric breakdown typically progresses through streamer/leader development toward a conducting arc column, with rapid Joule heating and subsequent hydrodynamic expansion shaping the MIR radiance. Classical gas-discharge and spark-discharge frameworks provide a useful qualitative picture for these stages and their characteristic time scales.\cite{Raizer1991GasDischarge,Bazelyan1998SparkDischarge} The arc emission contains strong broadband thermal radiation and line emission from the high-temperature plasma, making it a representative hazardous transient for engineering surveillance. In this perspective, the relevant question is not only whether the event can be detected, but also whether its \emph{approach-to-arc} regime admits decision variables that are more reliable than a single intensity threshold under realistic drift and background fluctuations.

\begin{figure}[t]
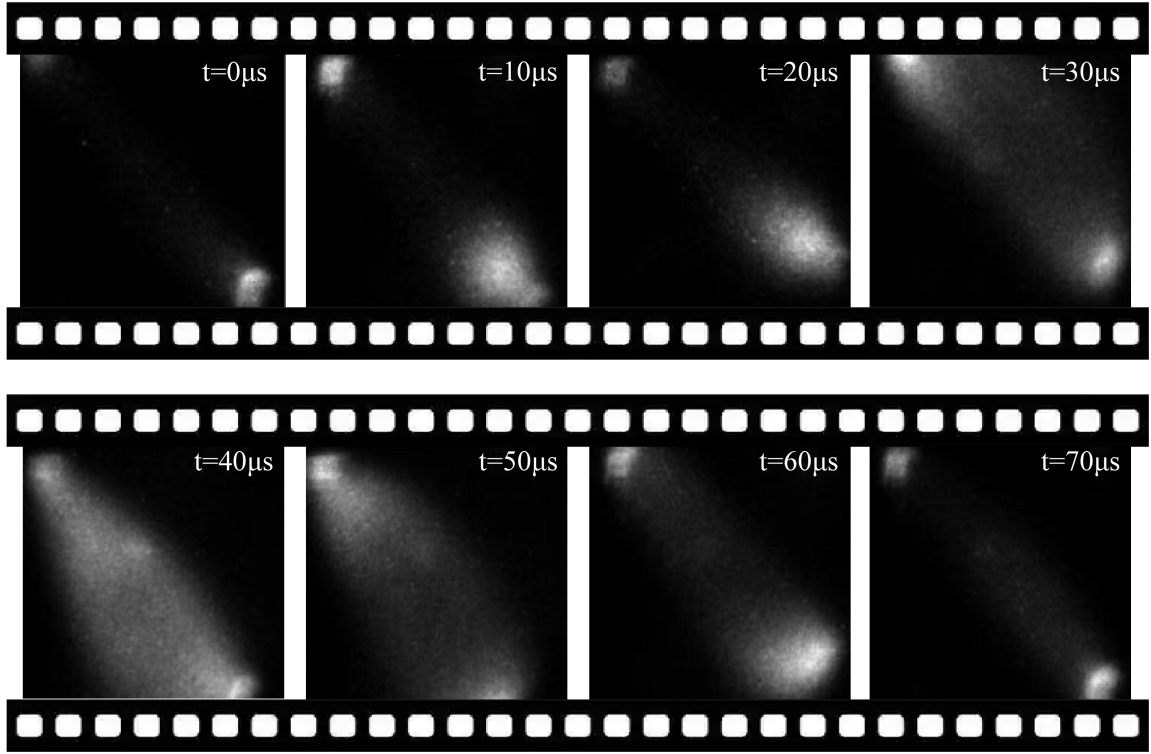

  \centering
  \incfig{\FigArcDemo}{width=0.92\linewidth}{Electric-arc image sequence or representative frames captured at 100\,kHz.}
  \caption{Ultrafast passive MIR upconversion imaging of an air-breakdown arc at $\mathrm{100\,kHz}$, capturing the onset and rapid evolution of the plasma/thermal emission.}
  \label{fig:arc_demo}
\end{figure}

Figure~\ref{fig:arc_demo} presents representative frames (or a time-ordered sequence) captured at 100\,kHz. The full evolution, including ignition, rapid expansion, and decay, is resolved. The iCCD is operated in fast-kinetics mode at a fixed physical frame rate of 100\,kHz, and a single (fixed) intensifier gate is chosen to capture both the weak onset emission and the subsequent bright phase without saturating the detector. To further ensure that VIS/NIR radiation does not enter the system, we switched off the pump laser and repeated the same acquisition; when the pump is off, no valid arc image is obtained.

A useful way to interpret such a hazardous transient is to view ignition as a rapid transition between two regimes: a weakly emitting pre-breakdown state and a high-radiance conducting column. In Fig.~\ref{fig:arc_demo}, the first $\sim$20~$\mu$s show weak, spatially localized emission, followed by a rapid rise and a pronounced expansion/migration of the bright region over $\sim$30--50~$\mu$s, consistent with conductive-channel establishment, intense Joule heating, and subsequent hydrodynamic expansion of the hot gas. From an engineering-surveillance standpoint, this naturally suggests a \emph{temperature-risk} framing: the key is to detect (and ideally predict) an impending over-temperature/plasma-column formation from short-horizon image statistics, rather than waiting for a saturated bright phase.

Past atom-based microwave sensing studies in our group offer two concrete methodological anchors for this goal.\cite{ding2022np,ding2024prl} First, the critical-metrology viewpoint emphasizes that sensitivity is governed by an effective ``susceptibility'' of the readout to the underlying state variable.\cite{ding2022np} Here the analogous quantity is the radiance-to-temperature response (or more generally the rate at which band-integrated radiance changes as the hot channel forms), which can be operationally accessed from the movie via a brightness proxy such as the ROI-summed signal $S_k$ (or $\mu_k$) and its time derivative. Even without absolute thermometry, one may define an \emph{equivalent} temperature indicator $T_k^\ast$ through a calibrated counts-to-radiance mapping, so that $dT^\ast/dt$ or $d\ln S/dt$ quantifies the onset growth rate of thermal loading. 

In the present system the imaging band is broadband, so $T_k^\ast$ should be interpreted as a radiometrically weighted indicator rather than a line-by-line thermodynamic temperature. Nevertheless, the same framework suggests a practical route to temperature-targeted warning: by inserting a narrowband filter (or a small set of selectable bands) that emphasizes the spectral region whose Planck radiance is most sensitive around a designated hazardous temperature, one can effectively ``tune'' the susceptibility of $S_k$ to that temperature range. In such a configuration, a rise of $T_k^\ast$ (or equivalently a persistent increase in $S_k$ relative to baseline) provides a direct, bandwidth-engineered precursor metric, enabling threshold-based alarms when the scene approaches and crosses the specified temperature level, while retaining the microsecond-resolved temporal readout for fast transients.

Second, the early-warning/tipping-point perspective highlights that ``symptom'' regimes can emerge in fluctuation statistics before the macroscopic jump.\cite{ding2024prl} Translating this to arc surveillance, the pre-bright phase can be characterized by short-window statistics extracted from $\mu_k$ and $\Delta_k$ (Eqs.~\eqref{eq:metric_mean}--\eqref{eq:metric_diff}), such as a rising variance within a fixed ROI, increased temporal autocorrelation (critical-slowing-down-like behavior), or a systematic change in spatial moments/bright-area growth rate. These quantities provide decision variables that are more robust than a single-frame intensity threshold, while remaining compatible with the drift-aware stability Allan deviation analysis reported next.

Beyond demonstrating ``that it can be seen'', the sequence in Fig.~\ref{fig:arc_demo} reveals dynamical features that are difficult to access with conventional MIR cameras at lower frame rates. In the earliest frames the emission is weak and spatially localized, consistent with the onset of ionization and the formation of a conductive channel. As the discharge develops, the radiance rises abruptly and the bright region expands and/or migrates, reflecting rapid Joule heating, plasma density growth, and the establishment of a fully developed arc column. The high-brightness core and the lower-brightness halo can evolve differently: the core typically tracks the hottest plasma column while the halo reflects rapidly expanding heated gas and collection effects. At later times, the emission region may deform and show apparent motion, consistent with the combined influence of electrode geometry, buoyancy-driven flow, and stochasticity in the discharge path. Capturing these precursors and the transition into the bright phase provides access to growth rates, expansion velocities, and intermittency statistics from the image sequence.

Moreover, the availability of a spatiotemporal movie enables decision variables that are more robust than a single-frame intensity threshold. For example, features derived from $\mu_k$ and $\Delta_k$ (Eqs.~\eqref{eq:metric_mean}--\eqref{eq:metric_diff}) such as $d\mu/dt$, the temporal evolution of spatial moments, or the rate of change of the bright-region area can help discriminate genuine onset from benign fluctuations. In the spirit of early-warning analyses near tipping points, one may also track whether fluctuation measures (e.g., variance within a preselected ROI, or the dispersion of $\mu_k$ across repeated shots) deviate from their baseline scaling as the system approaches the bright phase, thereby providing a ``symptom'' regime preceding the fully developed arc.\cite{ding2024prl} Interpreting such short-horizon statistics, however, requires a drift-aware understanding of measurement stability, motivating the Allan-deviation-based calibration reported next.

\FloatBarrier

\subsection{Allan-deviation calibration}
\label{sec:results_allan}

This subsection quantifies drift-aware stability limits of the ultrafast imaging system under controlled illumination. We first introduce a compact noise taxonomy and the Allan-deviation formalism, and then apply it to the ROI metrics used throughout the paper to obtain operationally meaningful bounds on averaging and stability.

For passive imaging, the measured ROI time series contains both fast stochastic noise and slow drift. We model the measured metric $x_k$ (either $\mu_k$ or $\Delta_k$) as
\begin{equation}
\label{eq:noise_model}
 x_k = s_k + b_k + m_k s_k + n_k,
\end{equation}
where $s_k$ is the true signal component, $b_k$ denotes slow baseline drift (e.g., thermal drift, camera bias), $m_k$ is a multiplicative fluctuation (e.g., pump-power jitter that scales the converted signal), and $n_k$ is additive stochastic noise (e.g., read noise, background photon noise). Standard time-averaging reduces $n_k$ but is ineffective against drift $b_k$, and can even amplify the impact of $m_k$ for long integration. Allan deviation provides a practical route to quantify the optimal averaging time in the presence of drift, and has been widely adopted for stability characterization in precision optical measurements and camera-based instrumentation.\cite{Czerwinski2009OE,Ossenkopf2008stability} For a uniformly sampled sequence $x_k$ with sampling interval $\tau_0$, the (overlapping) Allan deviation at averaging time $\tau=m\tau_0$ is
\begin{equation}
\label{eq:allan}
\sigma_x^2(\tau) = \frac{1}{2}\left\langle\left(\bar{x}_{i+1}^{(m)}-\bar{x}_{i}^{(m)}\right)^2\right\rangle,
\end{equation}
where $\bar{x}_{i}^{(m)}$ is the average of $m$ consecutive samples in bin $i$. The Allan deviation $\sigma_x(\tau)$ typically decreases as $\tau^{-1/2}$ when white noise dominates, reaches a minimum, and then increases when drift or low-frequency noise dominates; the minimizer $\tau^*$ offers an operationally meaningful integration/averaging time.

\begin{figure}[t]
  \centering
  \includegraphics[width=0.92\linewidth]{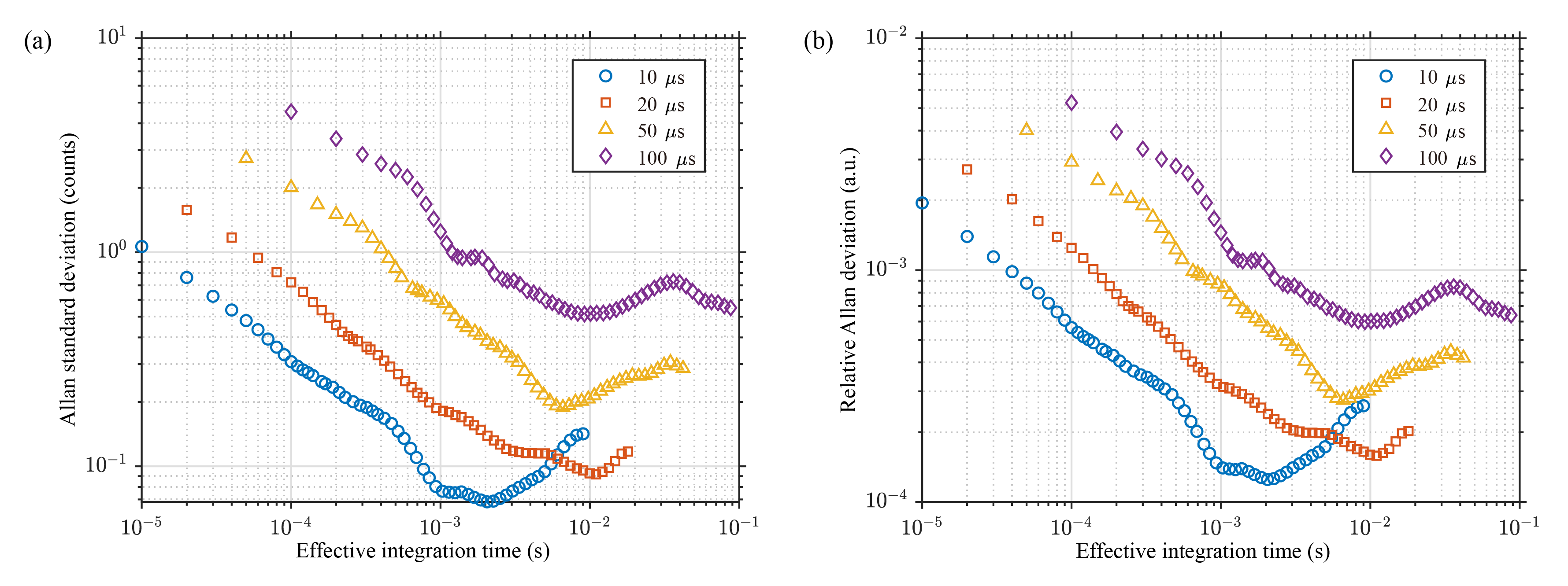}
  \caption{Allan deviation analysis for the ROI mean counts under four representative gate widths $\tau_g$. (a) Allan standard deviation. (b) Relative Allan deviation.}
  \label{fig:allan_mean}
\end{figure}

We apply Allan analysis to time series acquired under controlled tungsten-lamp illumination (Fig.~\ref{fig:setup_lamp}). We report two complementary metrics: (i) the ROI mean $\mu$ (Eq.~\eqref{eq:metric_mean}), representing an absolute measurement, and (ii) the bright--dark differential $\Delta$ (Eq.~\eqref{eq:metric_diff}), representing a drift-suppressed measurement. Figure~\ref{fig:allan_mean} summarizes Allan results for the ROI mean $\mu$ across multiple gate widths $\tau_g$ (four representative settings are shown). In Fig.~\ref{fig:allan_mean}(a), at short averaging times, $\sigma_\mu(\tau)$ decreases approximately as $\tau^{-1/2}$, consistent with a regime dominated by uncorrelated noise (shot noise from background/scene photons and camera readout noise). As $\tau$ increases, the curves reach a minimum that marks the crossover where further averaging no longer yields net benefit because low-frequency components (baseline drift $b_k$ and multiplicative fluctuations $m_k$ in Eq.~\eqref{eq:noise_model}) begin to dominate; the long-$\tau$ upturn is therefore an operational signature that the measurement becomes drift-limited.

Two gate-dependent trends are particularly informative. In the absolute Allan-deviation view (Fig.~\ref{fig:allan_mean}(a)), longer gates typically yield larger absolute fluctuations simply because the mean signal level is higher, so the four curves do not collapse even when they share a similar noise mechanism. In the relative/uncertainty view (Fig.~\ref{fig:allan_mean}(b)), this trivial scaling is removed, and the remaining spread reflects genuine differences in stability: longer gates improve photon statistics (lower relative white-noise floor) but can become more sensitive to slow gain drifts and pump-induced multiplicative noise, which manifests as an earlier or stronger long-$\tau$ upturn. The gate-dependent non-overlap in the drift-limited regime is thus a concrete experimental fingerprint that absolute mean measurements are jointly limited by additive offsets and multiplicative gain variations. Practically, the minimum of $\sigma_\mu(\tau)$ and its location provide a quantitative criterion for selecting an operating point: it gives the longest averaging time that remains beneficial before drift dominates, and its dependence on $\tau_g$ reveals how the gate choice trades photon statistics against drift susceptibility.

The Allan-deviation traces in Fig.~\ref{fig:allan_mean}(a,b) exhibit the canonical ``decrease--minimum--increase'' behavior expected for a system in which short-$\tau$ fluctuations are dominated by approximately white noise, while long-$\tau$ behavior is limited by slow drift and correlated noise. For an ROI-averaged mean signal $\mu$, the white-noise contribution scales approximately as $\sigma_{\rm A}(\tau)\propto \tau^{-1/2}$, reflecting the reduction of uncorrelated fluctuations by averaging, whereas slow drift yields an upturn that can be described phenomenologically as $\sigma_{\rm A}(\tau)\propto \tau^{\alpha}$ with $\alpha>0$ over the relevant time range. The minimum therefore identifies an optimal effective integration time $\tau^{\star}$ at which the transition between these noise regimes occurs. Importantly, varying the intensifier gate width $\tau_g$ changes the per-frame photon budget and the relative weight of shot/read noise versus drift, so the short-$\tau$ separation of the four curves mainly reflects gate-dependent white-noise levels, while the convergence of their minima indicates a common drift timescale set by the upconversion optics, pump-power fluctuations, and camera bias stability.

\begin{figure}[t]
  \centering
  \includegraphics[width=0.92\linewidth]{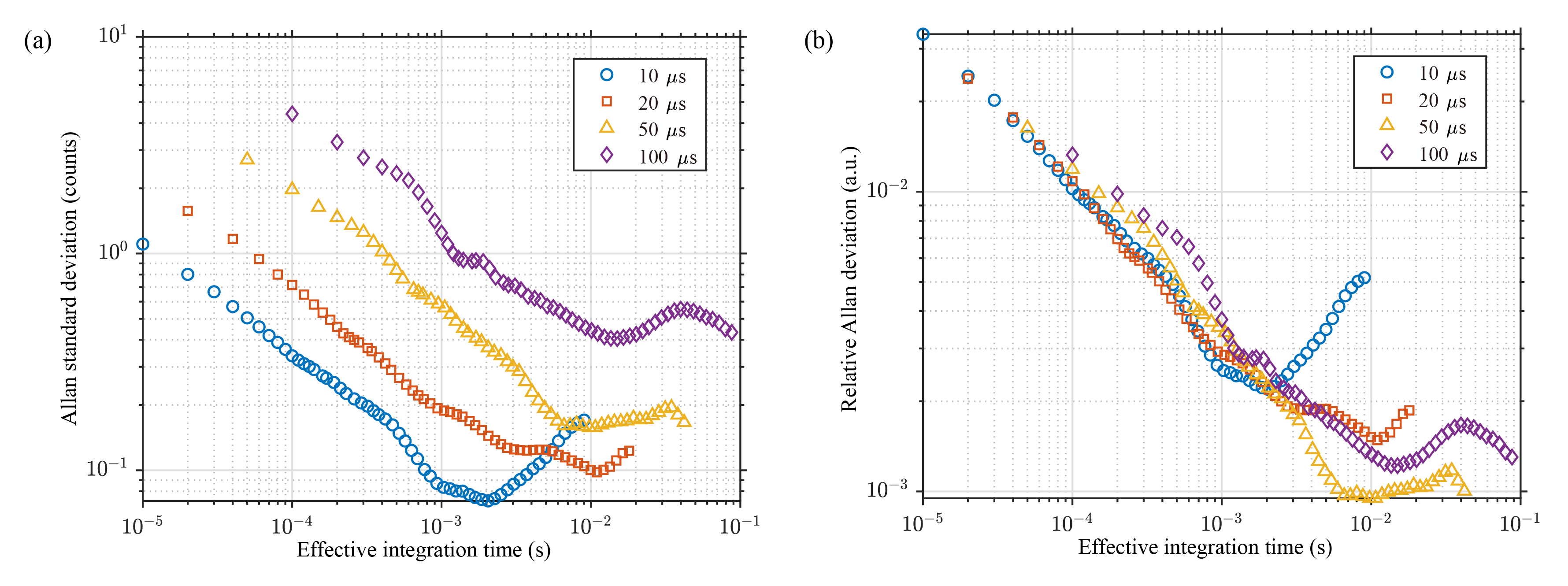}
  \caption{Allan deviation analysis for the bright--dark differential counts under four representative gate widths $\tau_g$. (a) Allan standrad deviation. (b) Relative Allan deviation.}
  \label{fig:allan_diff}
\end{figure}

Figure~\ref{fig:allan_diff} reports Allan results for the differential metric $\Delta$, again for multiple gate widths (four representative settings are shown). In contrast to the absolute mean, subtraction of a nearby dark/background region removes a large portion of additive offsets and common-mode drift, so the residual time series is closer to a ``true'' signal fluctuation plus uncorrelated noise. This change is visible in two ways. First, the short-$\tau$ region still follows the expected $\tau^{-1/2}$ scaling, indicating that stochastic noise is averaged down in the same manner. Second, the long-$\tau$ upturn is delayed and the curves collapse more strongly across different $\tau_g$, i.e., the overlap between gate settings improves, especially in the relative/uncertainty representation (Fig.~\ref{fig:allan_diff}(b)) where trivial signal scaling is removed.

The overlap carries physical meaning. Strong overlap implies that the dominant slow term is common-mode between the bright and dark regions and is therefore rejected by differencing---consistent with additive background drift and camera bias being the primary long-term limitation. Conversely, any residual separation between the curves points to mechanisms that are not perfectly common-mode, most notably multiplicative fluctuations that modulate only the converted signal path, imperfect background selection, or photon-starved operation at the shortest gate where white noise remains significant and the crossover occurs earlier. In our data, the differential metric produces a more consistent Allan minimum across gate widths, which directly simplifies engineering calibration.

Figure~\ref{fig:allan_diff}(a,b) further highlights why bright--dark differencing is advantageous for high-speed measurements. By computing $\Delta=\mu_{\rm bright}-\mu_{\rm dark}$ on the same frame sequence, additive backgrounds (pump leakage, stray light, camera offset, and slow common-mode drifts) are largely cancelled, so that the residual fluctuations are closer to a multiplicative-noise model dominated by pump-power jitter and gain variations. This manifests as a markedly improved overlap of the Allan-deviation curves across the four $\tau_g$ settings and a lower Allan floor, i.e., reduced sensitivity of the calibration to the particular choice of gate width. The remaining gate dependence is most visible for the shortest gate: the 10\,$\mu$s trace turns upward earlier because its reduced photon number per frame increases the white-noise coefficient, shifting the crossover to drift-limited behavior to smaller $\tau$, whereas the longer-gate traces share an almost identical minimum because their crossover is set primarily by the same low-frequency drift spectrum after common-mode rejection.

Taken together, these Allan signatures provide drift-aware bounds for selecting averaging horizons and for interpreting microsecond-resolved image statistics under realistic long-term stability constraints.

\section{Conclusion and outlook}
\label{sec:conclusion}

We demonstrated passive incoherent ultrafast MIR upconversion imaging at a physical frame rate of 100\,kHz using a CPLN-based SFG stage and a silicon iCCD operated in fast-kinetics mode. As a representative hazardous transient, we captured the evolution of an air-breakdown electric arc with microsecond time resolution, illustrating how passive MIR imaging can reveal rapid onset and morphology changes that are inaccessible to conventional MIR cameras at lower frame rates. Importantly, this capability aligns with deployment-oriented early-warning needs: in contrast to sensing paradigms that primarily confirm an event after it has occurred (as is common in UV-triggered monitoring), microsecond-resolved MIR movies provide access to precursor-like growth and expansion signatures, while remaining compatible with stand-off, passive surveillance enabled by atmospheric transmission windows.

A key message for practical use is that system performance is governed not only by frame rate but also by stability and drift. We therefore introduced Allan deviation analysis as a drift-aware calibration tool to quantify the crossover from white-noise-limited averaging to drift-limited behavior and to determine an operationally meaningful averaging horizon $\tau^*$ under realistic additive and multiplicative noise. In addition, differential ROI processing enhances robustness by suppressing baseline offsets and common-mode drift, leading to improved consistency across gate settings and simplifying the interpretation of ROI-based time series. Beyond one-time calibration, Allan deviation can also be evaluated during operation using a rolling buffer, providing compact indicators of measurement readiness through changes in the short-$\tau$ noise floor, shifts of $\tau^*$, and the emergence of long-$\tau$ upturns associated with pump fluctuations, thermal drift, or background variation.

Looking forward, the most immediate gains for field deployment will come from engineering refinements that widen the drift-tolerant operating window identified by Allan analysis, including improved pump stabilization, tighter thermal control, and more effective stray-light suppression. On the system side, extending the front-end optics to wider fields of view and longer stand-off configurations will enable practical thermal surveillance in open environments, where passive operation and robustness to uncontrolled backgrounds are essential. On the algorithmic side, burst-mode pipelines that perform on-the-fly feature extraction and sequential decision logic can translate microsecond-resolved images into actionable early-warning outputs, rather than single-frame alarms. Finally, spectrally selective or multi-band upconversion could help separate continuum thermal radiation from plasma line emission, improving the physical interpretability of observed onset dynamics and strengthening robustness across diverse transient scenarios.

\section*{Acknowledgements}
We acknowledge the support from the National Key Research and Development Program of China (2022YFB3903102),
National Natural Science Foundation of China (NSFC) (62435018),
Innovation Program for Quantum Science and Technology (2021ZD0301100),
USTC Research Funds of the Double First-Class Initiative (YD2030002023),
Quantum Science and Technology-National Science and Technology Major Project (2024ZD0300800),
and Research Cooperation Fund of SAST, CASC (SAST2022-075).
We also thank colleagues for helpful discussions.The authors declare no conflicts of interest.

\printbibliography

@article{Dam2012NP,
  title        = {Room-temperature mid-infrared single-photon spectral imaging},
  author       = {Dam, Jeppe Skovsgaard and Pedersen, Christian and Tidemand-Lichtenberg, Peter},
  journaltitle = {Nature Photonics},
  volume       = {6},
  number       = {11},
  pages        = {788--793},
  year         = {2012},
  doi          = {10.1038/nphoton.2012.231}
}

@article{Dam2012OE,
  title        = {Theory for upconversion of incoherent images},
  author       = {Dam, Jeppe Skovsgaard and Tidemand-Lichtenberg, Peter and Pedersen, Christian},
  journaltitle = {Optics Express},
  volume       = {20},
  number       = {2},
  pages        = {1475--1482},
  year         = {2012},
  doi          = {10.1364/OE.20.001475}
}

@article{Dam2010OL,
  title        = {High-resolution two-dimensional image upconversion of incoherent light},
  author       = {Dam, Jeppe Skovsgaard and Tidemand-Lichtenberg, Peter and Pedersen, Christian},
  journaltitle = {Optics Letters},
  volume       = {35},
  number       = {22},
  pages        = {3796--3798},
  year         = {2010},
  doi          = {10.1364/OL.35.003796}
}

@article{Junaid2019Optica,
  title        = {Video-rate, mid-infrared hyperspectral upconversion imaging},
  author       = {S. Junaid and S. Chaitanya Kumar and M. Mathez and M. Hermes and N. Stone and N. Shepherd and M. Ebrahim-Zadeh and P. Tidemand-Lichtenberg and C. Pedersen},
  journaltitle = {Optica},
  volume       = {6},
  number       = {6},
  pages        = {702--708},
  year         = {2019},
  doi          = {10.1364/OPTICA.6.000702}
}

@article{Huang2022NatCommun,
  title        = {Wide-field mid-infrared single-photon upconversion imaging},
  author       = {Huang, Kun and Fang, Jianan and Yan, Ming and Wu, E and Zeng, Heping},
  journaltitle = {Nature Communications},
  volume       = {13},
  pages        = {1077},
  year         = {2022},
  doi          = {10.1038/s41467-022-28716-8}
}

@article{Fang2023NatCommun,
  title        = {High-speed scanless entire bandwidth mid-infrared chemical imaging},
  author       = {Zhao, Yue  and  Kusama, Shota  and  Furutani, Yuji  and  Huang, Wei Hong  and  Luo, Chih Wei  and  Fuji, Takao },
  journaltitle = {Nature Communications},
  volume       = {14},
  pages        = {3929},
  year         = {2023},
  doi          = {10.1038/s41467-023-39628-6}
}

@article{Ge2023PRApplied,
  author       = {Ge, Zheng and Han, Zhaoqizhi and Liu, Yiyang and Wang, Xiaohua and Zhou, Zhiyuan and Yang, Fan and Li, Yinhai and Li, Yan and Chen, Li and Li, Wuzhen and Niu, Sujian and Shi, Baosen},
  title        = {Midinfrared up-conversion imaging under different illumination conditions},
  journaltitle = {Phys. Rev. Appl.},
  volume       = {20},
  number       = {5},
  pages        = {054060},
  year         = {2023},
  doi          = {10.1103/PhysRevApplied.20.054060},
}

@article{Ge2023APN_NETD,
  title={Thermal camera based on frequency upconversion and its noise-equivalent temperature difference characterization},
  author={Ge, Zheng and Zhou, Zhi-Yuan and Cheng, Jing-Xin and Chen, Li and Li, Yin-Hai and Li, Yan and Niu, Su-Jian and Shi, Bao-Sen},
  journal={Advanced Photonics Nexus},
  volume={2},
  number={4},
  pages={046002--046002},
  year={2023},
}

@article{Tamamitsu2020Optica,
author = {Tamamitsu, Miu and Toda, Keiichiro and Shimada, Hiroyuki and Honda, Takaaki and Takarada, Masaharu and Okabe, Kohki and Nagashima, Yu and Horisaki, Ryoichi and Ideguchi, Takuro},
year = {2020},
month = {04},
pages = {359-366},
title = {Label-free biochemical quantitative phase imaging with mid-infrared photothermal effect},
volume = {7},
journal = {Optica},
doi = {10.1364/OPTICA.390186}
}

@article{Paiva2022AnalChem,
author = {Paiva, Eduardo M.  and  Schmidt, Florian M.},
year = {2022},
month = {10},
pages = {14242-14250},
title = {Ultrafast Widefield Mid-Infrared Photothermal Heterodyne Imaging},
volume = {94},
  number       = {41},
journal = {Analytical Chemistry},
doi = {10.1021/acs.analchem.2c02548}
}

@article{Kviatkovsky2020SciAdv,
  title        = {Microscopy with undetected photons in the mid-infrared},
  author       = {Kviatkovsky, Ilia and Chrzanowski, Helen M. and Lipfert, Malte and others},
  journaltitle = {Science Advances},
  volume       = {6},
  number       = {42},
  pages        = {eabd0264},
  year         = {2020},
  doi          = {10.1126/sciadv.abd0264}
}

@article{Lahiri2015PRA,
  title        = {Theory of quantum imaging with undetected photons},
  author       = {Lahiri, Mayukh and Lapkiewicz, Radek and Lemos, Gabriel and Zeilinger, Anton},
  journaltitle = {Physical Review A},
  volume       = {92},
  number       = {1},
  pages        = {013832},
  year         = {2015},
  doi          = {10.1103/PhysRevA.92.013832}
}

@article{Ossenkopf2008stability,
  title={The stability of spectroscopic instruments: a unified Allan variance computation scheme},
  author={Ossenkopf, Volker},
  journal={Astronomy \& Astrophysics},
  volume={479},
  number={3},
  pages={915--926},
  year={2008},
  publisher={EDP Sciences},
  doi          = {10.1051/0004-6361:20079188}
}

@article{Czerwinski2009OE,
  title        = {Quantifying noise in optical tweezers by allan variance},
  author       = {Fabian and Czerwinski and  Andrew, C  and Richardson and  Lene, B  and Oddershede},
  journaltitle = {Optics Express},
  volume       = {17},
  number       = {15},
  pages        = {13255--13269},
  year         = {2009},
  doi          = {10.1364/OE.17.013255}
}

@article{ding2022np,
  title   = {Enhanced metrology at the critical point of a many-body Rydberg atomic system},
  author  = {Ding, Dong-Sheng and Liu, Zong-Kai and Shi, Bao-Sen and Guo, Guang-Can and M{\o}lmer, Klaus and Adams, Charles S.},
  journal = {Nature Physics},
  volume  = {18},
  number  = {12},
  pages   = {1447--1452},
  year    = {2022},
  doi     = {10.1038/s41567-022-01777-8}
}

@article{ding2024prl,
  title   = {Early Warning Signals of the Tipping Point in Strongly Interacting Rydberg Atoms},
  author  = {Zhang, Jun and Zhang, Li-Hua and Liu, Bang and Zhang, Zheng-Yuan and Shao, Shi-Yao and Li, Qing and Chen, Han-Chao and Liu, Zong-Kai and Ma, Yu and Han, Tian-Yu and Wang, Qi-Feng and Adams, C. Stuart and Shi, Bao-Sen and Ding, Dong-Sheng},
  journal = {Physical Review Letters},
  volume  = {133},
  number  = {24},
  pages   = {243601},
  year    = {2024},
  doi     = {10.1103/PhysRevLett.133.243601}
}

@book{Raizer1991GasDischarge,
  title     = {Gas Discharge Physics},
  author    = {Raizer, Y. P.},
  year      = {1991},
  publisher = {Springer},
  address   = {Berlin}
}

@book{Bazelyan1998SparkDischarge,
  title     = {Spark Discharge},
  author    = {Bazelyan, E. M. and Raizer, Y. P.},
  year      = {1998},
  publisher = {CRC Press},
  address   = {Boca Raton}
}

@article{Park2024ArcFlashUV,
  title   = {Evaluation of Time-Based Arc Flash Detection with Non-contact UV Sensor},
  author  = {Park, Chulmin and Lee, Kiwon and Kim, Kinam and Lim, Hunyoung and Park, Young},
  journal = {Journal of Electrical Engineering \& Technology},
  year    = {2024},
  volume  = {19},
  pages   = {1983--1992},
  doi     = {10.1007/s42835-023-01555-3}
}

@article{Riba2022UVImagingDischargeReview,
  title   = {Application of Image Sensors to Detect and Locate Electrical Discharges: A Review},
  author  = {Riba, Jordi-Roger},
  journal = {Sensors},
  year    = {2022},
  volume  = {22},
  number  = {15},
  pages   = {5886},
  doi     = {10.3390/s22155886}
}

@article{Shen2023TransportableMIRLHR,
  title   = {Performance Characterization of a Fully Transportable Mid-Infrared Laser Heterodyne Radiometer (LHR)},
  author  = {Shen, Fengjiao and Hu, Xueyou and Lu, Jun and Xue, Zhengyue and Li, Jun and Tan, Tu and Cao, Zhensong and Gao, Xiaoming and Chen, Weidong},
  journal = {Sensors},
  year    = {2023},
  volume  = {23},
  number  = {2},
  pages   = {978},
  doi     = {10.3390/s23020978}
}
\end{document}